\documentclass[aps,prl,reprint,groupedaddress]{revtex4-2}
\usepackage{bm}
\usepackage{amsmath}
\usepackage{graphicx}
\usepackage{dcolumn}%
\usepackage{color}
\usepackage{comment}

\begin{document}

\title{Effective Temperature and Einstein Relation for Particles in Mesoscale Turbulence}

\author{Sanjay CP and Ashwin Joy}
\email[]{ashwin@physics.iitm.ac.in}
\affiliation{Department of Physics, Indian Institute of Technology Madras, Chennai - 600036}

\date{\today}

\begin{abstract}
  From the smallest scales of quantum systems to the largest scales of intergalactic medium, turbulence is ubiquitous in nature. Often dubbed as the last unsolved problem of classical physics, it remains a time tested paradigm of dynamics far from equilibrium. The phenomenon even transcends to self-propelled fluids such as dense bacterial suspensions that can display turbulence at mesoscale even though the constituent particles move at Reynolds number below unity.   It is intensely debated whether such fluids possess an effective temperature and obey fluctuation-dissipation relations (FDR) as they are generally marred by a lack of detailed balance. In this letter, we answer this question and report an exact expression of the effective temperature for a distribution of interacting particles that are advected by a mesoscale turbulent flow. This effective temperature is linear in particle diffusivity with the slope defining the particle mobility that is higher when the background fluid exhibits global polar ordering, and lower when the fluid is in isotropic equilibrium. We believe our work is a direct verification of the Einstein relation \textemdash the simplest FDR, for interacting particles immersed in a mesoscale turbulence.
\end{abstract}


\maketitle

Fluid turbulence has been fascinating to us since time immemorial. From the deluge drawings of Leonardo da Vinci \cite{doi:10.1146/annurev-fluid-022620-122816} dating almost five centuries ago to the relatively modern works on complex fluids \cite{RevModPhys.80.225}, plasmas \cite{Yamada:2008uc} and atmospheric systems \cite{Wyngaard:1992th}, fluid turbulence is ubiquitous in nature. Despite many decades of exhaustive research, this chaotic state of fluid motion continues to remain as the oldest unsolved problem in classical physics. Put simply, there is currently no rigorous theory that starts from the Navier-Stokes equation and leads to the experimentally observed laws on energy dissipation and velocity structure functions in turbulent flows \cite{frisch1995turbulence,davidson2015turbulence}. Nevertheless, turbulence is considered as a classical paradigm of nonlinear dynamics far from thermodynamic equilibrium \cite{Ruelle20344,Goldenfeld:2017ve,cardy_falkovich_gawedzki_2008}. An exciting area of mesoscale  turbulence has recently emerged where the fluid displays spatio-temporal chaos even though its constituent units are self-propelled and move at Reynolds number below unity. Examples include dense bacterial suspensions \cite{PhysRevLett.93.098103,PhysRevLett.107.028102}, mircotubule networks \cite{Sanchez:2012vi}, artificial swimmers  \cite{PhysRevE.92.052309} and active liquid crystals \cite{Alert:2020tx}, to mention a few. It is a matter of intense debate whether these self-propelled fluids should satisfy fluctuation-dissipation relations (FDR) as the constituent particle dynamics generally lack detailed balance \cite{sym13010081,PhysRevE.103.032607}. A related and important question is whether one can describe an effective temperature of particles dispersed in such fluids \cite{Cugliandolo_2011,PUGLISI20171}. In this report, we address these fundamental issues by predicting an exact expression of the effective temperature for a distribution of interacting particles that are advected by an active turbulent flow in the background. This effective temperature is linear in particle diffusivity with the slope characterizing the particle mobility, that is higher when the background fluid exhibits global polar ordering, and lower when the fluid is in isotropic equilibrium. Our work is therefore a direct verification of the celebrated Einstein relation - the simplest FDR, for interacting particles advected in mesoscale turbulence. The results reported here are valid across four decades of variation in the damping coefficient, thereby putting a large number of active systems within the ambit of our work \textemdash from dense suspensions of microswimmers that are traditionally overdamped \cite{Gompper_2020,RevModPhys.85.1143,RevModPhys.88.045006} to the lesser understood underdamped suspensions where inertial effects are significant, notable examples being ciliates, planktons, copepods and other macroswimmers moving in a background media of low viscosity \cite{C9SM01019J}. In what follows, we provide the details of the phenomenological model and methods used in our work.

\textit{Model/Methods:} To mimic a generic active suspension, we use a minimal continuum model of an incompressible active fluid in two dimensions \cite{wensink2012meso, dunkel2013minimal}, whose streamfunction $\Psi(x,y)$ evolves according to the following equation
\begin{eqnarray}
  \frac{\partial}{\partial t}(\nabla^2 \Psi) + \lambda_0 \frac{\partial (\Psi, \nabla^2 \Psi)}{\partial(x,y)} = -\Gamma_0 \nabla^4 \Psi - \Gamma_2 \nabla^6 \Psi - \mu \nabla^2 \Psi\nonumber\\
  \label{active_model}
\end{eqnarray}
here the non-dimensional parameter $\lambda_{0}$ decides the type
of active unit, meaning they are either \textit{pusher} or a
\textit{puller}, corresponding respectively to the case
$\lambda_{0} > 0$ or $\lambda_{0} < 0$. The model was used
recently to investigate pattern formation
\cite{PhysRevFluids.3.061101} and transport
coefficients in dense active liquids
\cite{PhysRevFluids.5.024302}. We set the value of $\lambda_0 =
3.5$ throughout our work implying that we have a \textit{pusher}
type of active units. We keep $\Gamma_{0,2} > 0$ to enable energy
injection into the active liquid via fluid instabilities. The
scalar field $\mu = \alpha + \beta |\bm u|^2$ depends on the
local velocity $\bm u = \bm \nabla \times \Psi \hat{z}$, and was
first introduced by Toner and Tu  to model the ``flocking''
behavior in self-propelled rod-like objects
\cite{toner2005hydrodynamics,toner1998flocks}. The parameter
$\alpha$, henceforth referred to as the Ekman friction, acts at
intermediate scales and can either lead to a damping of energy
when $\alpha > 0$ or an injection of energy when $\alpha <0$.
Former leads the fluid to an isotropic equilibrium and the latter
yields  a globally ordered polar state with mean velocity
$\sqrt{|\alpha|/\beta}$. To non-dimensionalize eq.
(\ref{active_model}), we normalize all distances to $\sigma_0 =
5\pi\sqrt{2\Gamma_2 / \Gamma_0}$ and all times to $t_0 =
5\pi\sqrt{2}\Gamma_2/\Gamma_0^2$. In terms of these reduced
units, we fix the values of the model parameters as $\Gamma_0 =
(5\pi\sqrt{2})^{-1}, \Gamma_2 = (5\pi\sqrt{2})^{-3}$ and $\beta =
0.5$, in order to remain consistent with earlier works
\cite{bratanov2015new,PhysRevFluids.5.024302}. Equation (\ref{active_model}) is then numerically solved using a pseudo-spectral approach over a square grid of $512^{2}$ points in a doubly periodic box of size $2\pi$ \cite{canuto2012spectral}. A time step of $\Delta t = 2 \times 10^{-4}$ is employed  for time marching $\Psi$ with the Crank-Nicholson scheme that is sufficient to maintain numerical stability in the entire range of parameters explored here. 

Into the active turbulent flow, we throw $N$ particles that follow the dynamics:
\begin{eqnarray}
  \frac{\text{d}\bm r_i}{\text{d}t} &=& \bm v_i \nonumber\\
  \frac{\text{d}\bm v_i}{\text{d}t} &=& - \gamma  (\bm v_i - \bm u(\bm r_i)) - \bm \nabla_i \Phi 
  \label{Eq-Non-Markov}
\end{eqnarray}
where $\gamma$ is the damping coefficient and $(\bm r_i, \bm v_i)$ correspond respectively to position and velocity of an $i^{\text{th}}$ particle. We realize $\bm u(\bm r_i)$ as the turbulent flow field projected at the location of this particle, and  $-\bm \nabla_i \Phi$ as the mass normalized force acting on it. The limit $\gamma \longrightarrow \infty$ naturally corresponds to an overdamped dynamics of passive tracers advected by a background flow. In this limit, the tracers can exhibit an intervening  anamalous diffusive regime as the Ekman friction is pushed to negatively larger values, say $\alpha < -6$ \cite{mukherjee2021anomalous}. The limit $\gamma \longrightarrow 0$ on the other hand, corresponds to the Newtonian dynamics of particles without any background fluid. At intermediate values of $\gamma$, both inertial and damping effects coexist thereby rendering generality to the particle dynamics. We set the interaction between particles as the repulsive Weeks-Chandler-Andersen potential energy $\Phi = \sum_{j<k} \phi(|\bm r_j - \bm r_k|)$, where the potential
\begin{equation}
  \phi (r) = \begin{cases} 4 \epsilon \bigg[\bigg(\dfrac{\sigma_0}{r}\bigg)^{12} - \bigg(\dfrac{\sigma_0}{r}\bigg)^{6}\bigg] + \epsilon, & r < 2^{1/6} \sigma_0 \\
    0, & r > 2^{1/6} \sigma_0 
  \end{cases}
\end{equation}
All measurements on the particles as well as the turbulent fluid are carried out only after the latter attains a steady state. To improve our statistics, we average these measurements over 100 independently prepared realizations of the initial state. Below we show that the statistics of the Eulerian velocity $\bm u$ is indeed profitable to the prediction of an effective temperature of  our particles. 

\begin{figure}[htp]
  \centering
  \includegraphics[width=0.5\textwidth,keepaspectratio]{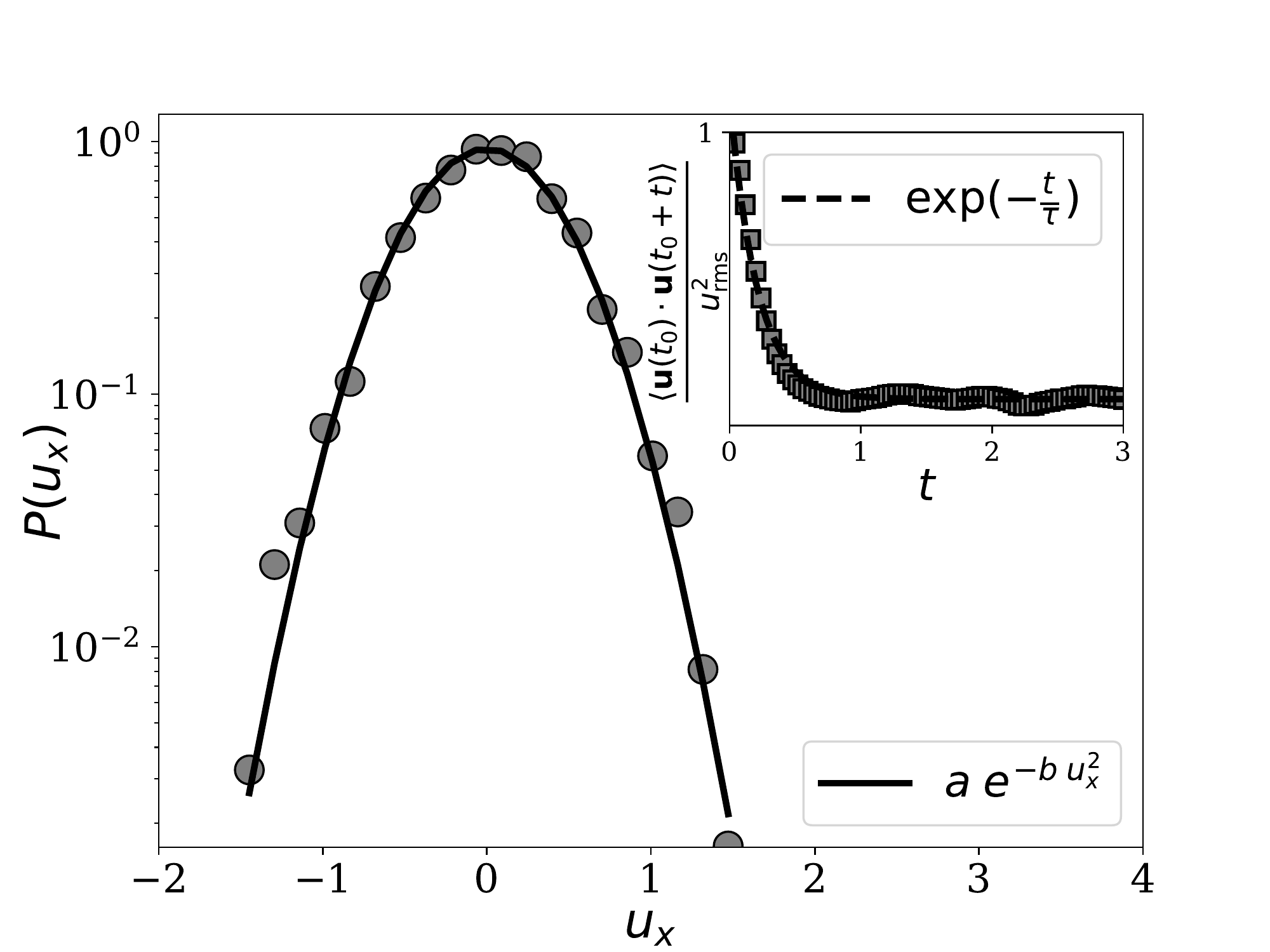}
  \caption{Distribution of Eulerian velocities projected on the particles fits well to a Gaussian. Data shown here corresponds to $\alpha = 3$; results are similar for other $\alpha$. Inset: Temporal auto-correlation of $\bm u$ decays exponentially with a relaxation time $\tau$. We can thus visualize $\bm u$ as a Gaussian colored noise - a headway that will be used to formulate an effective temperature for our particles.}
  \label{Fig-pu}
\end{figure}

\textit{Statistics of $\bm u$:} We start with a measurement of the Eulerian flow field projected at the particle positions, i.e $\bm u(\bm r_i)$. In Fig. (\ref{Fig-pu}), we show the distribution of a component of this velocity in the steady state and realize that it is well approximated by a Gaussian. In the inset, we show its temporal auto-correlation that is clearly an exponential with a relaxation time $\tau$. Combining these two facts, we can claim that the fluid velocity projected at particle locations is a Gaussian colored noise of zero mean and variance given as -
\begin{equation}
  \langle u_{i\chi}(t) u_{j\psi}(t')\rangle = (1/2) \delta_{ij} \delta_{\chi \psi} u^2_{\text{rms}} \text{exp}(-|t-t'|/\tau)  \label{Eq-ucorr}
\end{equation}
where $u_{\text{rms}}$ is a velocity scale that sets the kinetic energy of the background fluid in the steady state. We use the Latin symbols $i, j$ to denote particle labels and the Greek symbols $\chi, \psi$ to denote spatial indices. In the light of all these arguments, we realize that the stochastic dynamics in Eq. (\ref{Eq-Non-Markov}) is essentially non-Markovian and its corresponding steady state Fokker-Planck equation must be carefully solved (see next) to derive an exact expression of the stationary probability distribution. Factorization of this distribution into position and momentum parts will yield an exact expression of the effective temperature of our particles. This is done next.

\begin{figure*}[htp]
\centering
\begin{tabular}{cc}
  \includegraphics[width=.5\textwidth]{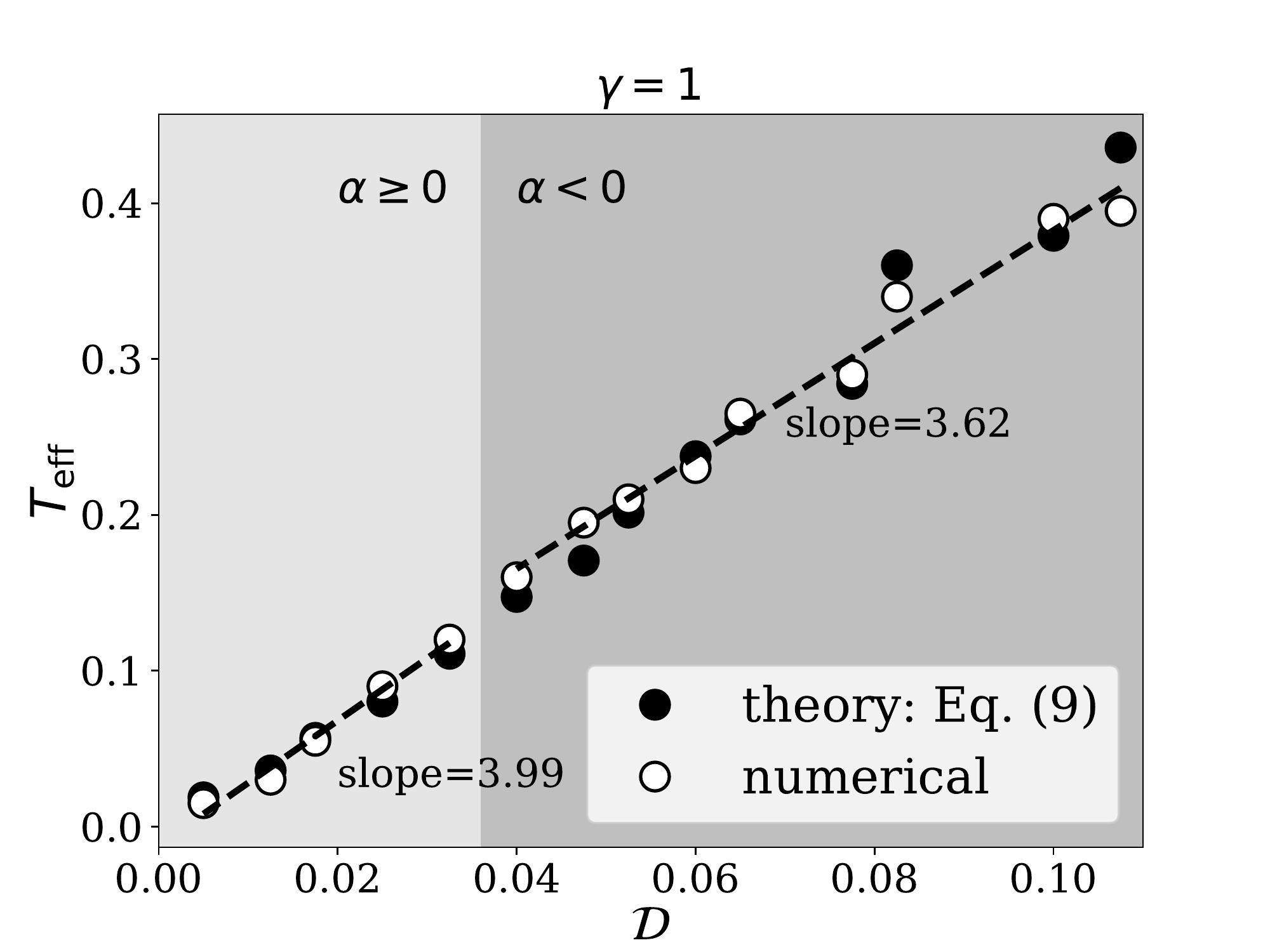}
  &
  \includegraphics[width=.5\textwidth]{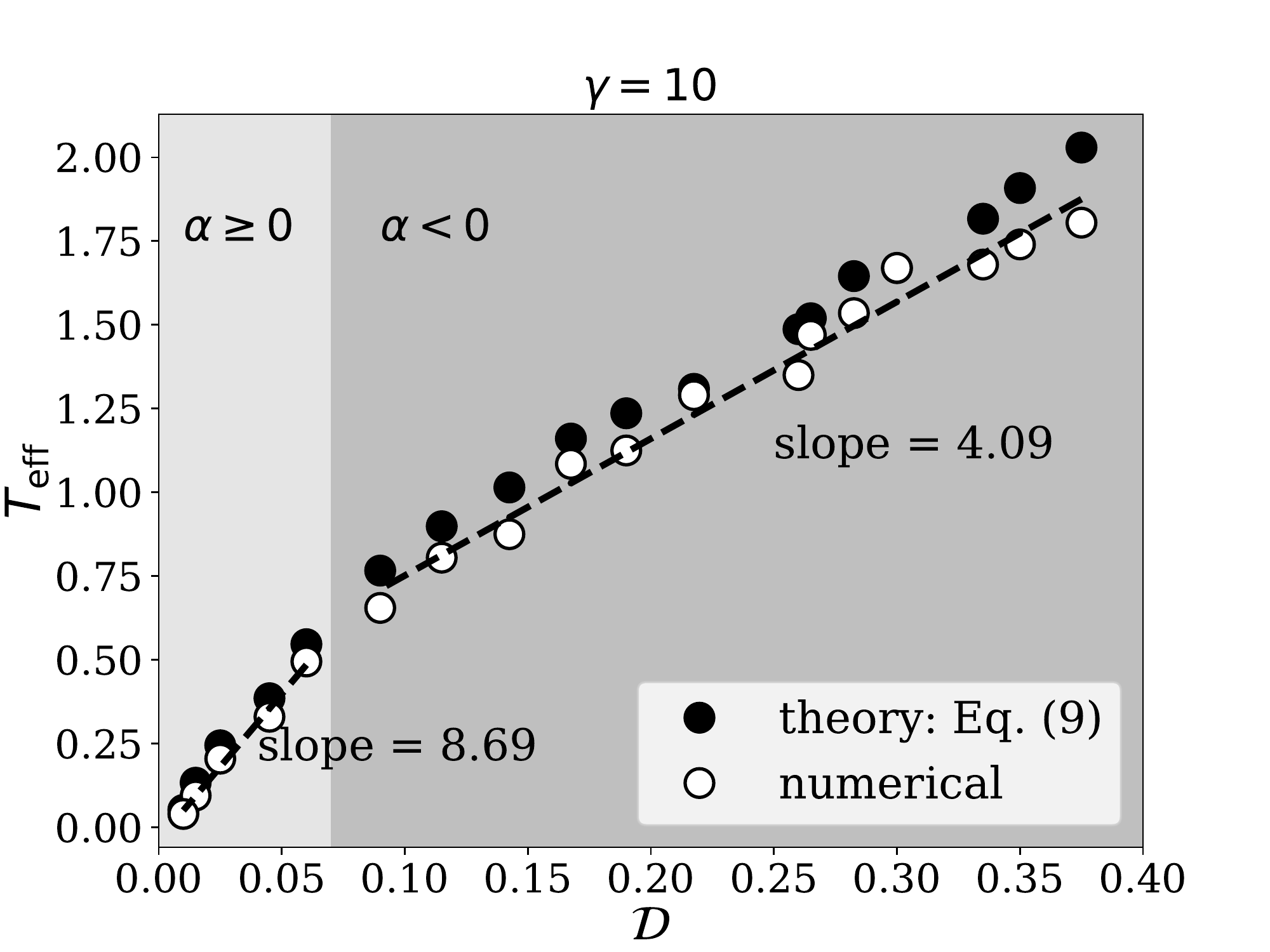}\\
  \includegraphics[width=.5\textwidth]{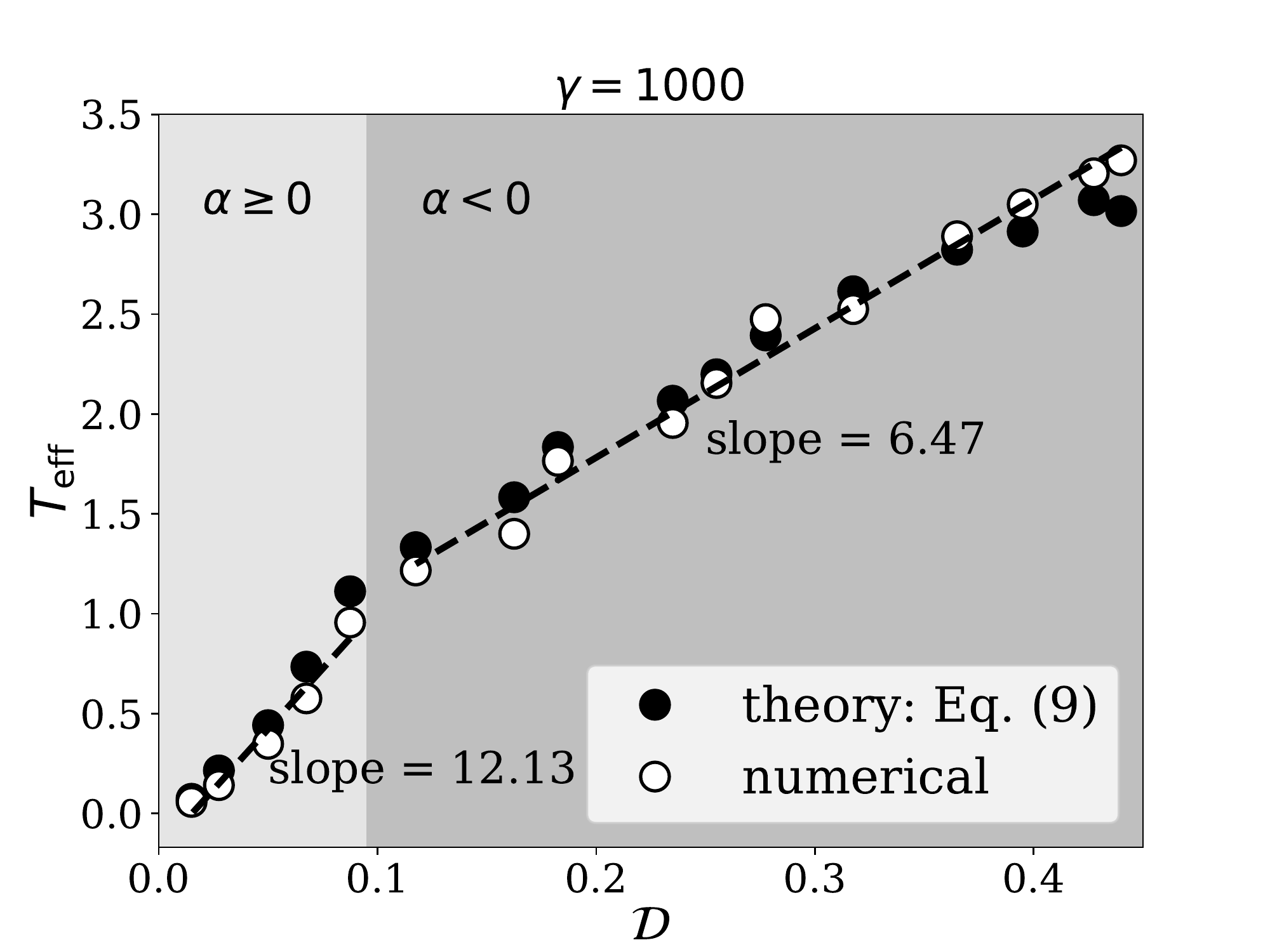}
  &
  \includegraphics[width=.5\textwidth]{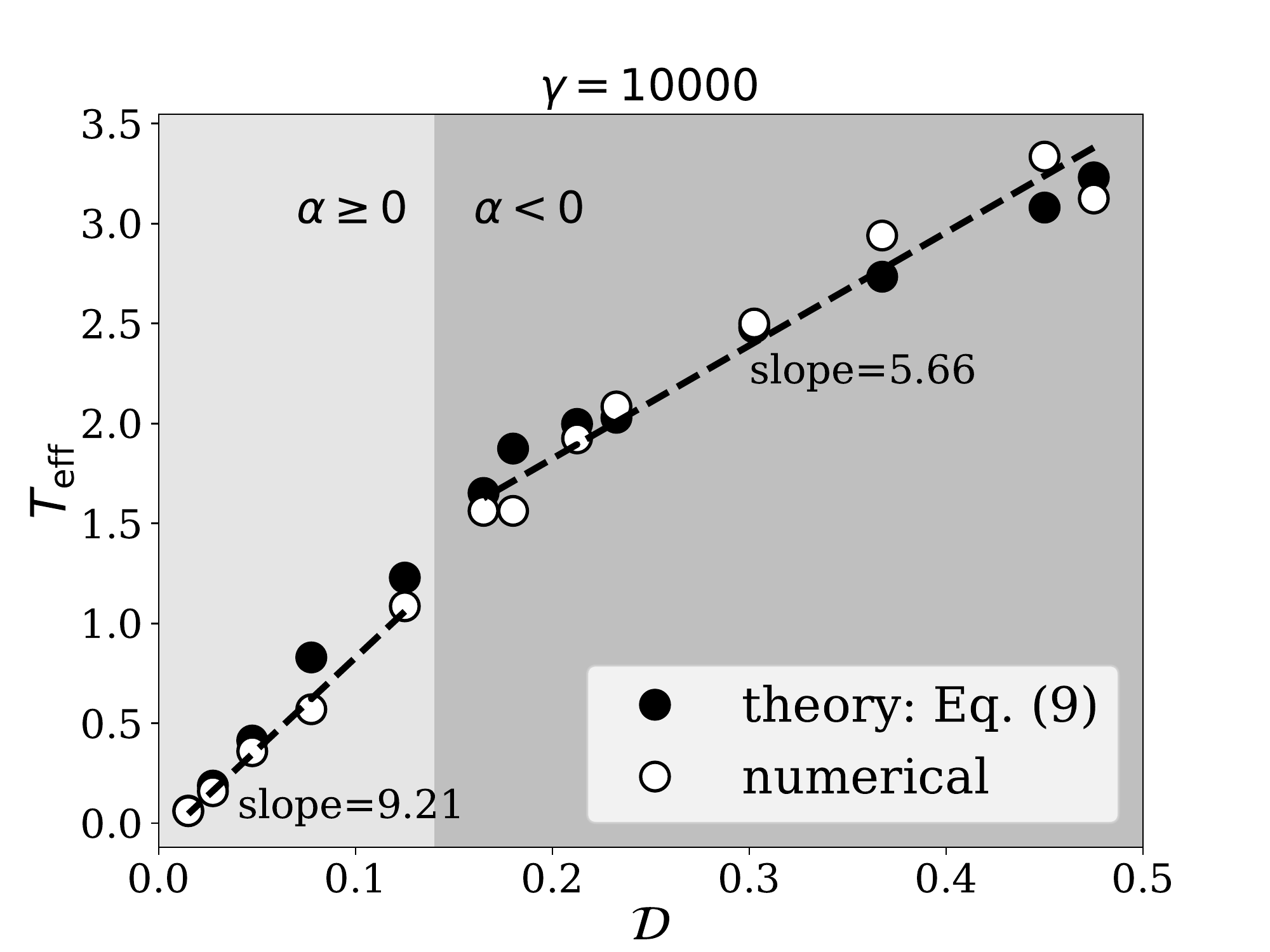}
\end{tabular}
\caption{Evidence of linear relationship between effective temperature $T_{\text{eff}}$ and particle diffusivity $D$ is presented here at various value of damping coefficient $\gamma$. The numerical value of $T_{\text{eff}}$ is extracted by fitting $\bm u$ to a Gaussian and the theoretical value is predicted by Eq. (\ref{Eq-Teff}). Dashed line is a linear fit to our numerical data and has a slope that indicates the inverse of particle mobility $\mu$. This is a direct verification of the Einstein's relation, the simplest FDR. Particle diffusivity $D=\lim_{t \to \infty} \langle \Delta r^2 \rangle/(4t)$ is varied by varying the fluid friction $\alpha$. For each panel that corresponds to a specific $\gamma$, the mobility of particles is higher when the background fluid is in a globally ordered polar state ($\alpha < 0$), and lower when it is in isotropic equilibrium ($\alpha > 0$).}
\label{Fig-Eintein-Law}
\end{figure*}

\textit{Theoretical Prediction of} $T_{\text{eff}}$: To systematically derive an expression for the effective temperature $T_{\text{eff}}$ in terms of the fluid properties, we will now set up a Fokker-Planck equation for the governing dynamics of the interacting particles and solve it to get a stationary velocity distribution. As stated earlier, the dynamics shown in Eq. (\ref{Eq-Non-Markov}) is essentially non-Markovian due to the Gaussian colored noise $\bm u$. An exact solution of the corresponding Fokker-Planck equation, with the exception of harmonic interactions, is intractable in the general case and approximations are always required \cite{1995AdChP..89..239H}. To circumvent this difficulty, we reduce the original non-Markovian dynamics in Eq. (\ref{Eq-Non-Markov}) to an equivalent Markovian dynamics by an appropriate enlargement of the phase space \cite{van1992stochastic,risken1996fokker}. In the component form, this simply reads 
\begin{eqnarray}
  \frac{\text{d} r_{i\chi}}{\text{d}t} &=& v_{i\chi} \nonumber\\
  \frac{\text{d} v_{i\chi}}{\text{d}t} &=& -\gamma (v_{i\chi} - u_{i\chi}) - \nabla_{i\chi} \Phi \nonumber\\
\frac{\text{d} u_{i\chi}}{\text{d}t} &=& -\frac{1}{\tau} u_{i\chi} + \sqrt{\frac{u^2_{\text{rms}}}{\tau}} \xi_{i\chi}
 \label{Markov}
\end{eqnarray}
where $\bm \xi_i$ is a Gaussian white noise with zero mean and correlation $\langle \xi_{i\chi}(t) \xi_{j\psi}(t') \rangle = \delta_{ij} \delta_{\chi \psi} \delta (t-t')$. The reader should note that the dynamics in Eq. (\ref{Markov}) is consistent with the expected correlation properties of $\bm u$ that we already know from Eq. (\ref{Eq-ucorr}). The Fokker-Planck equation for the probability distribution $\mathcal{P}(r_\chi, v_\chi, t)$ corresponding to the dynamics in (\ref{Markov}) can be written using a decoupling approximation as \cite{PhysRevA.33.3320}:
\begin{equation}
  \frac{\partial}{\partial t} \mathcal{P}(r_\chi, v_\chi, t) = \mathcal{L} \;\mathcal{P}(r_\chi, v_\chi, t)
  \label{FP}
\end{equation}
  where the operator $\mathcal{L}$ is defined as 
\begin{widetext}
\begin{equation}
  \mathcal{L} = \bigg[-v_{\chi}  \frac{\partial}{\partial r_\chi} + \frac{\partial}{\partial v_\chi}\bigg(\gamma v_\chi + \frac{\partial \Phi}{\partial r_\chi}\bigg) + \dfrac{\gamma^2 u^2_{\text{rms}}\tau}{2\bigg(1 + \gamma \tau + \tau^2 \bigg\langle \dfrac{\partial^2 \Phi}{\partial r_\chi^2} \bigg\rangle\bigg)} \dfrac{\partial^2}{\partial v^2_\chi} + \dfrac{\gamma^2 u^2_{\text{rms}}\tau^2}{2\bigg(1 + \gamma \tau + \tau^2 \bigg\langle \dfrac{\partial^2 \Phi}{\partial r_\chi^2} \bigg\rangle\bigg)} \dfrac{\partial^2}{\partial x_\chi \partial v_\chi}\bigg],
  \label{FP}
\end{equation}
\end{widetext}
The steady state solution of the above Fokker-Planck equation readily factorizes into position and velocity parts as:
\begin{equation}
  \mathcal{P}_{\text{s}} (r_\chi, v_\chi) = \frac{1}{\mathcal{Z}}\exp\bigg(-\frac{v^2_\chi}{2 \sigma^2_\chi}\bigg) \exp\bigg[-\frac{\Phi}{\sigma^2_\chi (1 + \gamma \tau)}\bigg]
  \label{P_st}
\end{equation}
with the velocity variance realized as  
\begin{equation}
  \sigma^2_\chi = \frac{u^2_{\text{rms}} \gamma \tau}{2\bigg(1 + \gamma \tau + \tau^2 \bigg\langle \dfrac{\partial^2 \Phi}{\partial x^2_\chi}\bigg\rangle \bigg)} \equiv T_{\text{eff}} 
  \label{Eq-Teff}
\end{equation}
the effective temperature of our inertial particles, with $\chi$ being $x$ or $y$. We take a moment of pause to appreciate the beauty of this exact expression. There are no fitting parameters here and we are able to completely predict the effective temperature of particles once the background fluid properties ($u_{\text{rms}}, \tau, \gamma$) and particle interactions ($\Phi$) are specified. The predicted $T_{\text{eff}}$ is in excellent agreement with the numerically estimated value that is obtained by fitting the velocity distribution to a Gaussian. This is shown in Fig. \ref{Fig-Eintein-Law} where we plot $T_{\text{eff}}$ vs. $D$, the particle diffusivity. For each $\gamma$, we calculate the diffusivity as $D=\lim_{t \to \infty} \langle \Delta r^2 \rangle/(4t)$ and its variation $D_{\text{min}} \rightarrow D_{\text{max}}$ comes from varying the fluid friction $\alpha$ in the range $4.0\rightarrow-4.0$. It is evident from each panel of this figure that $T_{\text{eff}}$ is linear in $D$ with a slope yielding the inverse of particle mobility $\mu$. This is a direct verification of the Einstein's relation, the simplest formulation of FDR. We also observe that the particle mobility is higher when the background fluid is in a globally ordered polar state, i.e $\alpha < 0$, and lower when the fluid is in isotropic equilibrium, or $\alpha > 0$. Our results are valid across four decades of variation in damping coefficient effectively covering  the entire range $1 < \gamma\tau < 10^4$, where the lower and upper limit corresponding respectively to the inertial and overdamped regime. This puts a large number of active systems within the purview of our work\textemdash from overdamped suspensions of microswimmers \cite{wu2000particle,PhysRevLett.110.198302}  to the underdamped suspensions of macroswimmers such as ciliates, planktons, copepods moving in low viscosity media \cite{C9SM01019J}.

\textit{Summary:} We have demonstrated that a distribution of interacting particles immersed in an active turbulent fluid can be described by an effective temperature. To achieve this, we solved the steady state Fokker-Planck equation corresponding to the stochastic governing equations of the particles. The result is an exact relation between the effective temperature, properties of the background flow and particle interactions. Finally we show this effective temperature is linear in particle diffusivity with the slope characterizing particle mobility. This is a direct verification of the Einstein's relation - the simplest FDR, for particles advected by mesoscale turbulence. The mobility is observed to be higher when the background fluid is in a globally ordered polar phase, and lower when the fluid is in isotropic equilibrium. The results reported here apply to active liquids prepared under a broad spectrum of particle damping, and are therefore general in nature.

\begin{acknowledgements}
  The authors thank V Balakrishnan, Sumesh Thampi and Abhijit Sen for discussions and comments on the manuscript. All simulations were done on the HPC-Physics cluster of our group. Support from the SERB core grant, DST and New Faculty Seed Grant (NFSG), IIT Madras, is gratefully acknowledged.
\end{acknowledgements}

\bibliography{manuscript}

\end{document}